\documentstyle[epsfig]{article}
\begin{document}

\renewcommand{\thesection}{\arabic{section}.}
\renewcommand{\figurename}{\small Fig.}
\renewcommand{\theequation}{\arabic{section}.\arabic{equation}}

\newcommand{\eqreset}{\setcounter{equation}{0}}

\newcommand{\lsim}{
\,\raisebox{0.5mm}{$\scriptstyle <$}
\hspace{-2.1mm}\raisebox{-0.7mm}{$\scriptstyle\sim$}\,
}     

\newcommand{\gsim}{
\,\raisebox{0.5mm}{$\scriptstyle >$}
\hspace{-2.1mm}\raisebox{-0.7mm}{$\scriptstyle\sim$}\,
}

\begin{flushleft}

{\small\it Journal of Statistical Physics, Vol. 83, Nos. 5/6, 907--931, 1996 }
\vspace{3cm}

{
\Large\bf
The {\boldmath 1/$D$} Expansion for Classical Magnets: \\
Low-dimensional Models with Magnetic Field
}
\vspace{0.5cm}

\renewcommand{\thefootnote}{\fnsymbol{footnote}}
{\bf D. A. Garanin} 
\footnote[2]{E-mail: garanin@physnet.uni-hamburg.de}

{
I. Institut f\"ur Theoretische Physik, Universit\"at Hamburg,
Jungiusstr. 9, D-20355 Hamburg, Germany
}

\end{flushleft}

\begin{flushright}
\begin{minipage}[t]{10cm}
\small
{\it Received May 2, 1995} \\
\_\hrulefill \\
The field-dependent magnetization $m(H,T)$ of 1- and 2-dimensional classical 
magnets described by the $D$-component vector model is calculated 
analytically 
in the whole range of temperature and magnetic fields with the help of 
the $1/D$ expansion. In the \mbox{1-st} order in $1/D$ the theory reproduces 
with a good accuracy the temperature dependence of the zero-field 
susceptibility of antiferromagnets $\chi$ with the maximum 
at $T \lsim |J_0|/D$
($J_0$ is the Fourier component of the exchange interaction)
and 
describes for the first time the singular behavior of $\chi(H,T)$ at small 
temperatures and magnetic fields: 
$\lim_{T\to 0}\lim_{H\to 0} \chi(H,T)=1/(2|J_0|)(1-1/D)$ and 
$\lim_{H\to 0}\lim_{T\to 0} \chi(H,T)=1/(2|J_0|)$. \\
\_\hrulefill \\
KEY WORDS: Low-dimensional magnets; magnetic susceptibility; spherical model;
{\it 1/D} ({\it 1/n}) expansion.
\end{minipage}
\end{flushright}

\section{Introduction}
\eqreset

A great variety of low-dimensional magnetic systems were synthesized and 
experimentally investigated during the last  decades 
(see, e.g., refs. \cite {jon74} and \cite{man91}). The idealized 1- and 
2-dimensional models 
(without interplane or interchain coupling and anisotropy) are 
characterized by a 
strong short-range order in the low-temperature region, whereas the 
long-range order is ruled out being smeared off by the longwavelength spin 
waves. 
Complementary to the high-temperature series expansions (HTSE, 
see, e.g., refs.~\cite{rwood5863,lam75,cam75}), 
such approaches as the "modified spin wave theory" \cite{tak8789} and 
the "Schwinger boson mean field theory" \cite{aue88} were applied to 
low-dimensional antiferromagnets at low temperatures.  These 
two methods giving very similar results (with a wrong factor in the 
antiferromagnetic susceptibility in ref. \cite{aue88}) are not rigorous 
expansions in the parameter $T/|J|<<1$ ($J$ is the exchange integral)
but rather some variational 
approaches. The results break down, however, at $T\sim |J|$ and thus cannot 
describe the situation in the whole temperature range. Since the 
absence of magnetization was introduced in ref. \cite{tak8789} as  
some additional self-consistency requirement, the generalization for 
the case with the external magnetic field {\bf H} is a problem.

In ref. \cite{garst94} an analytical method of calculation 
of the physical quantities of {\it classical}\/ low-dimensional magnets 
in the whole temperature range was proposed, which is based on the $1/D$ 
expansion for the model of $D$-component classical spin vectors on a 
lattice \cite{sta6871}:
\begin{equation}\label{1.1}
{\cal H} = -{\bf H}\sum_{i}{\bf m}_i -
 \frac{1}{2}\sum_{ij}J_{ij}{\bf m}_i{\bf m}_j    \qquad |{\bf m}|=1
\end{equation}
with  the  help  of  the diagram 
technique developed in ref. \cite{garlut8486}. For the Heisenberg model 
($D=3$) in the \mbox{1-st} order in $1/D$ the calculated temperature dependences of 
the antiferromagnetic susceptibility and internal energy at $H=0$ 
turn out to be very good, which is shown by the 
comparison with the MC simulation data \cite{she80} for the internal 
energy of the square lattice (s.l.) classical ferromagnet and 
with the exact results \cite{sta69} for a "toy" example of the classical
linear chain (l.c.) 
model. In particular, for both models the characteristic maximum of 
the antiferromagnetic susceptibility at $T\lsim |J_0|/D$ is in contrast to 
refs. \cite{tak8789} and \cite{aue88} well reproduced. The reason for the 
efficiency of the $1/D$ expansion even for $D=3$ is that it 
yields the exact results for the thermodynamic 
quantities at $T\to 0$ and reproduces several leading terms of their 
HTSE expansion (see ref.~\cite{garst94}) interpolating thus between 
these limits in 
the whole temperature range. The applicability of the approach to 
classical low-dimensional magnets proposed in ref. \cite{garst94} is not 
restricted to the case $H=0$, and it can be applied to the
problem of the singular behavior of the antiferromagnetic susceptibility 
$\chi^{AF}(H,T)$ at low temperatures and magnetic fields, i.e. 
the inpermutability of its limits $\lim_{H\to 0}\lim_{T\to 0}$ and 
$\lim_{T\to 0}\lim_{H\to 0}$,  
which could not be up to now described by other analytical 
methods. The physical reason for such a singular behavior is the 
following. With lowering temperature the system becomes locally ordered,
and $D-1$ susceptibilities transverse with respect to the local
sublattice orientation tend to the value $1/(2|J_0|)$ 
($J_0=zJ$, $z$ is the number of nearest neighbors), whereas the 
longitudinal one tends to zero. At $H=0$ there is no preference 
direction, and the susceptibility of the sample is given by the average 
over the local sublattice orientations, which results in $\lim_{T\to 
0}\lim_{H\to 0} \chi^{AF}(H,T)=1/(2|J_0|)(1-1/D)$. For the Heisenberg 
model the $D$-dependent factor makes up the well-known number 2/3.
To the contrary, for the 
arbitrary small $H\neq 0$ at sufficiently low $T$ the lowest-energy 
state with the sublattice magnetizations 
driven perpendicular to the field ${\bf H}$ and tilted in the 
direction of ${\bf H}$ is realised. In this state the susceptibility 
takes up its transverse value: $\lim_{H\to 0}\lim_{T\to 0}
\chi^{AF}(H,T)=1/(2|J_0|)$. A quantitative description of this
effect and the calculation of the magnetization $m(H,T)$ of 
low-dimensional classical antiferromagnets in the whole range of
temperatures and magnetic fields with the help of the $1/D$
expansion is the purpose of this work.

The following part of the article is organized as follows. In Section 2 
an improved version of the diagrammatic $1/D$ expansion for classical 
spin systems with the magnetic field is developed and the results for 
the magnetization and spin-spin correlation function in the \mbox{1-st} order 
in $1/D$ are obtained. In Section 3 the general $1/D$-results, which are 
double integrals over the Brillouin zone, are calculated analytically 
and analyzed for the classical linear chain model, for which there is no 
exact solution in the case with the magnetic field. In Section 4 the 
results are converted into the form convenient for numerical 
calculations and the analysis at low temperatures for 2-dimensional 
systems, and the temperature  and magnetic field dependences of the 
antiferromagnetic susceptibility are represented. In Section 5 some 
important features of the $1/D$ expansion and its applicability to the 
systems with concrete values of $D$ are discussed.

\section{The 1/D expansion}
\eqreset

The physical quantities of ferro- and antiferromagnets described by 
the hamiltonian (\ref{1.1}) can be expanded in powers of $1/D$ with the 
help of the diagram technique for classical spin systems 
\cite{garst94,garlut8486}. 
Here the consideration of ref. \cite{garst94} is improved and generalized 
to the case $H\neq 0$. We choose the $z$ axis along the magnetic 
field ${\bf H}$, the other (transverse) components of the spin vector 
${\bf m}$ are designated by the index $\alpha=1,2,..,D-1$. The 
wavevector-dependent transverse susceptibility 
$\chi_\perp({\bf k})\equiv\chi_\alpha({\bf k})$ of a classical spin 
system is related to the
Fourier-transform of the spin-spin correlation function 
$S_{\alpha\alpha}({\bf r}-{\bf r}') =
<\!\!m_\alpha({\bf r})m_\alpha({\bf r}')\!\!>$ 
via the formula 
$\chi_\perp({\bf k})=\beta S_{\alpha\alpha}({\bf k})$, $\beta=1/T$. 
With the help of the diagram technique of ref. 
\cite{garst94} $\chi_\perp({\bf k})$ can be represented as
\begin{equation}\label{2.1}
\chi_\perp({\bf k})=
\frac{\beta \hat\Lambda_{\alpha\alpha}({\bf k})}
{1-\hat\Lambda_{\alpha\alpha}({\bf k})\beta J_{\bf k}}
\end{equation}
where $\hat\Lambda_{\alpha\alpha}({\bf k})$ is the compact (irreducible)
part of $S_{\alpha\alpha}({\bf k})$ given by the diagrams, which 
cannot be cut by the one interaction line $\beta J_{\bf k}$. For the 
isotrope systems considered here it is not necessary to write down 
the diagrams for the magnetization $m=<\!m_z\!\!>$, because $m(H)$ can be 
determined from (\ref{2.1}). Indeed, in a transverse magnetic field 
$H_\perp <<H$ the magnetization ${\bf m}$ is simply rotated on the 
angle $\theta=H_\perp/H<<1$, which results in the important relation 
\begin{equation}\label{2.2}
\chi_\perp\equiv \chi_\perp(0)=m/H.
\end{equation}
The longitudinal susceptibility can be determined now by the formula
\begin{equation}\label{2.3}
\chi_z \equiv \partial m/\partial H = \chi_\perp +
H(\partial \chi_\perp/\partial H)
\end{equation}
which is much easier than the direct 
diagrammatic calculation of $\chi_z({\bf k})$. 

The compact part $\hat\Lambda_{\alpha\alpha}({\bf k})$ in (\ref{2.1})
can be represented in the \mbox{1-st} order in $1/D$ by the diagram set from 
ref. \cite{garst94} completed by the additional diagrams for 
$H\not=0$, which 
can be estimated and selected according to the same rules. 
The general principle here is that the diagrams with multiple {\it 
irreducible} integrations over wavevectors (i.e. those that do not 
separate into products of independent simpler integrals) are small as 
the corresponding powers of $1/D$. Thus, in each order in $1/D$ the 
complexity of diagrams to be taken into account is restricted: in the 
zeroth order in $1/D$ (the spherical model) only diagrams with 
the one-loop integration over the Brillouin zone survive, 
and in the \mbox{1-st} order 
in $1/D$ the double integrals over the Brillouin zone appear. 
The large number of diagrams in the case $H\not=0$
necessitates, however, an improvement 
of the method, which consists in taking into account some diagrams {\it 
implicitly} with the subsequent solution of the corresponding equation 
for $\hat\Lambda_{\alpha\alpha}({\bf k})$. All the diagrams, 
which contribute to $\hat\Lambda_{\alpha\alpha}({\bf k})$ in the \mbox{1-st}
order in $1/D$, are represented in figs.1,2. 
Note that the 
renormalized transverse interaction lines in fig.1 contain the unknown 
quantity $\hat\Lambda_{\alpha\alpha}({\bf k})$ itself, which means
implicitly accounting for the additional class of diagrams of the 
type 1 and 2 in the fig.3 of ref. \cite{garst94}.
At $H=0$ from 
all the diagrams in fig.2,a survive only the diagrams 1, 2 and 5,
and the last term in the Dyson equation for the longitudinal interaction 
line fig.2,b disappears. The wavevector dependence of 
$\hat\Lambda_{\alpha\alpha}({\bf k})$ is due to the diagrams 5--8
in fig.2,a. There is one more diagram $7'$ that is analogous to
7 and is not represented in fig.2,a to save the place.
Taking into account only the diagrams in fig.1 results in the 
self-consistent Gaussian approximation (SCGA), which describes rather 
good the thermodynamics of 3-dimensional ferromagnets 
\cite {garlut8486}.  The analytical form of 
$\hat\Lambda_{\alpha\alpha}({\bf k})$ in fig.1 reads  
\begin{figure}[p]
\unitlength1cm
\begin{picture}(15,5)
\centerline{\epsfig{file=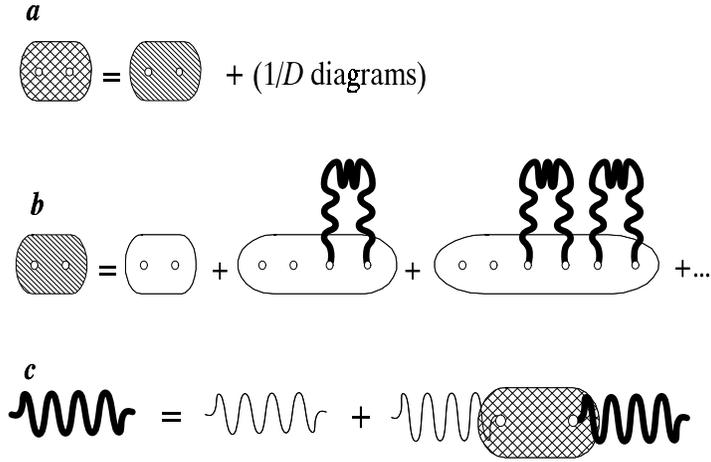,angle=0,width=16cm}}
\end{picture}
\par
\caption{  
(a) Diagrams for the compact part 
$\hat\Lambda_{\alpha\alpha}({\bf k})$ 
(see also fig.2,a);
(b) block summation of transverse loops for the renormalized cumulant one-site 
2-spin average $\tilde\Lambda_{\alpha\alpha}$;
(c) Dyson equation for the renormalized transverse interaction.
}
\end{figure}
\begin{figure}[p]
\unitlength1cm
\begin{picture}(15,7.4)
\centerline{\epsfig{file=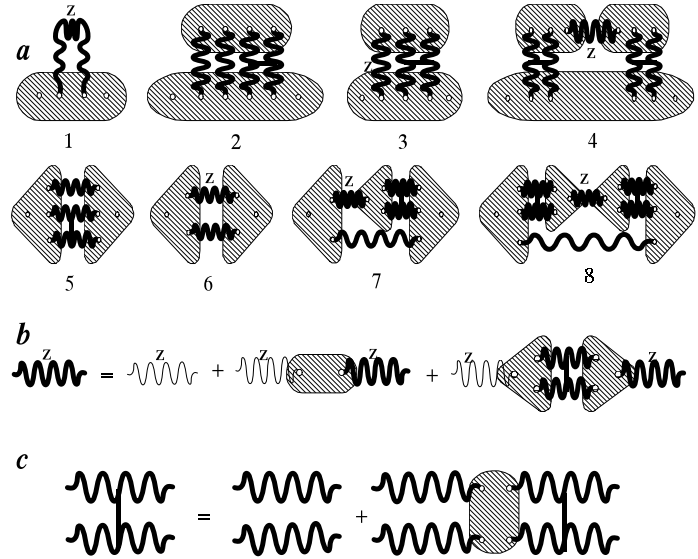,angle=0,width=20cm}}
\end{picture}\par
\caption{   
(a) Additional $1/D$-diagrams for 
$\hat\Lambda_{\alpha\alpha}({\bf k})$; 
(b) Dyson equation for the renormalized longitudinal interaction;
(c) ladder equation for the four-spin correlation line.
}
\end{figure}\par
\begin{equation}\label{2.4}
\hat\Lambda_{\alpha\alpha}({\bf k})=\tilde\Lambda_{\alpha\alpha}+
\hat\Lambda^{(1/D)}_{\alpha\alpha}({\bf k})
\end{equation}
where $\hat\Lambda^{(1/D)}_{\alpha\alpha}({\bf k})$ is the sum of the 
diagrams represented in fig.2,a vanishing in the limit $D\to\infty$ 
(see Appendix), and 
$\tilde\Lambda_{\alpha\alpha}$ is the renormalized 2-spin cumulant average 
given by \cite{garst94,garlut8486}
\begin{equation}\label{2.5}
\tilde\Lambda_{\alpha\alpha}=\frac{1}{\pi^{(D-1)/2}}\int\! d^{D-1}r\,
\exp(-r^2)\Lambda_{\alpha\alpha}(\mbox{\boldmath$\zeta$}).
\end{equation}
Here $\Lambda_{\alpha\alpha}$ is one of the cumulant spin averages of a 
general type \cite{garst94}
\begin{equation}\label{2.6}
\Lambda_{\alpha_1\alpha_2..\alpha_p}(\mbox{\boldmath$\xi$})=
\frac{\partial ^p\Lambda(\xi)}{\partial \xi_{\alpha_1} 
\partial \xi_{\alpha_2}..\partial \xi_{\alpha_p}}, 
\end{equation}
obtained through the generating function $\Lambda(\xi)=\ln Z_0(\xi)$,
\begin{equation}\label{2.7}
Z_0(\xi)={\rm const}\cdot \xi^{-(D/2-1)}{\rm I}_{D/2-1}(\xi)
\end{equation}
is the partition function of the $D$-component classical spin, ${\rm 
I}_\nu(\xi)$ is the modified Bessel function,
\begin{equation}\label{2.8}
\mbox{\boldmath$\zeta$} = \beta({\bf H+ m}J_0)+2l^{1/2}_\alpha{\bf r},
\end{equation}
${\bf r}$ is the $(D-1)$ -component vector perpendicular to ${\bf H}$. 
The last term in (\ref{2.8}) describes the Gaussian transverse 
fluctuations of the molecular field ${\bf H+ m}J_0$
with the dispersion proportional to $l_\alpha$, 
which leads to the renormalization of the cumulant spin averages 
described by (\ref{2.5}) for $\Lambda_{\alpha\alpha}$ and  by analogous 
formulas for the other cumulants entering 
$\hat\Lambda^{(1/D)}_{\alpha\alpha}({\bf k})$ (dashed ovals in figs.1,2). 
This renormalization results from the block summation of the all 
one-loop diagrams with the transverse interaction in fig.1,b; the 
quantity $l_\alpha$ is given by the integral over the Brillouin zone
\begin{equation}\label{2.9}
l_\alpha=\frac{1}{2}\,v_0\!\!\!\int\!\!\!\frac{d{\bf q}}{(2\pi)^d}
\frac{\beta J_{\bf q}}{1-\hat\Lambda_{\alpha\alpha}({\bf q})\beta 
J_{\bf q}}
\end{equation}
where $v_0$ is the unit cell volume ($v_0=a_0^d$ for the l.c. ($d=1$) 
and s.l. ($d=2$), $a_0$ is the atomic space). The $(D-1)$-dimensional 
integral in (\ref{2.5}) can be simplified taking advantage of the 
symmetry with respect to the transverse variables and using the 
explicite form
\begin{equation}\label{2.10}
\Lambda_{\alpha\alpha}(\mbox{\boldmath$\xi$})=
\frac{B(\xi)}{\xi}\left(1-\frac{\xi^2_\alpha}{\xi^2}\right)+
B'(\xi)\frac{\xi^2_\alpha}{\xi^2},
\end{equation}
where
\begin{equation}\label{2.11}
B(\xi) = \partial \Lambda(\xi)/\partial \xi = 
{\rm I}_{D/2}(\xi)/{\rm I}_{D/2-1}(\xi)
\end{equation}
is the generalized Langevin function and $B'(\xi)\equiv dB/d\xi$. Making in 
(\ref{2.10}) the substitution $\xi^2_\alpha=\xi^2_r/(D-1)$ with 
$\xi_r\equiv 2l^{1/2}_\alpha r$ and in (\ref{2.5}) the partial integration  to 
get rid of $B'$, one obtains
\begin{equation}\label{2.12}
\tilde\Lambda_{\alpha\alpha}=\frac{2}{\Gamma((D\!+\!1)\!/2)}
\int\limits^{\infty}_{0}\!dr\, r^D \exp(-r^2)\frac{B(\zeta)}{\zeta} \qquad 
\zeta=|\mbox{\boldmath$\zeta$}|
\end{equation}
The formulas (\ref{2.4}), (\ref{2.12}) and (\ref{2.9}) yield the 
integral equation for the compact part of the spin-spin correlation 
function $\hat\Lambda_{\alpha\alpha}({\bf k})$ entering the basic 
expression (\ref{2.1}).

By the expansion in powers of $1/D$ the quantities 
$\tilde\Lambda_{\alpha\alpha}$ and 
$\hat\Lambda^{(1/D)}_{\alpha\alpha}({\bf k})$ 
in (\ref{2.4}) give rise to the terms starting from the zero- and the 
\mbox{1-st} orders in $1/D$, correspondingly, whereas the expansion of all other 
diagrams 
neglected here starts from $1/D^2$. Before proceeding with the 
calculations we make a reference to the simplest approach --- the mean 
field approximation (MFA) --- in which no diagrams with the integration 
over wavevectors are taken into account. In this case 
$l_\alpha\Rightarrow 0$ and $\zeta=\xi=\beta(H+mJ_0)$, and in (\ref{2.4}) 
$\hat\Lambda^{(1/D)}_{\alpha\alpha}({\bf k}) \Rightarrow 0$,
$\tilde\Lambda_{\alpha\alpha} \Rightarrow \Lambda_{\alpha\alpha}
\Rightarrow B(\xi)/\xi$. Now with the use of (\ref{2.1}) and (\ref{2.2}) 
one gets the Curie-Weiss equation $m=B(\xi)$ for the magnetization $m$, 
which yields the phase transition temperature $T_C^{MFA}=|J_0|/D$. The 
latter has no physical significance for 1- and 2-dimensional magnets but 
can be used as a temperature scale. It is convenient to introduce the 
dimensionless temperature $\theta\equiv T/T_C^{MFA}$, magnetic field 
$h\equiv H/|J_0|$ and susceptibility $\tilde\chi\equiv |J_0|\chi$. Then 
the formulas (\ref{2.1}) and (\ref{2.9}) can be rewritten as
\begin{equation}\label{2.13}
\tilde\chi_\perp({\bf k}) =
 \frac{\hat G_{\bf k}}{1-\nu\hat G_{\bf k} 
\lambda_{\bf k}} \qquad
\tilde l_\alpha \equiv \frac{l_\alpha}{D} = 
\frac{1}{2\theta}\,v_0\!\!\!\int\!\!\!\frac{d{\bf q}}{(2\pi)^d}
\frac{\lambda_{\bf q}}{1-\nu\hat G_{\bf 
q}\lambda_{\bf q}}
\end{equation}
where $\hat G_{\bf k} \equiv 
(D/\theta)\hat\Lambda_{\alpha\alpha}({\bf k})$, 
$\nu=\pm 1$ for ferro- and antiferromagnets and 
$\lambda_{\bf k} \equiv J_{\bf k}/J_0$. 
In the integral (\ref{2.12}) the product $r^D\exp(-r^2)$ 
is at large $D$ sharply peaked at $r=r_0=(D/2)^{1/2}$, whereas 
$B(\zeta)/\zeta$ changes slowly with $r$. Using the expansion of $B(\xi)$ 
(\ref{2.11}) for $D\gg 1$ \cite{garst94} one can write
\begin{equation}\label{2.14}
\frac{B(\zeta)}{\zeta} \cong \frac{2}{D} 
\left(
g(x) + \frac{1}{D}
\frac{x^2}{1+x^2}g^2(x) 
\right); 
\qquad
g(x)=\frac{1}{1+(1+x^2)^{1/2}} 
\end{equation}
where
\begin{equation}\label{2.15}
x \equiv 2\zeta/D = x(\tilde r) =
\left(
4(h+\nu m)^2/\theta^2 + 8\tilde l_\alpha \tilde r^2 
\right)^{1/2} ;
\qquad
\tilde r \equiv r/r_0
\end{equation}
and evaluate (\ref{2.12}) by the pass method. In the \mbox{1-st} order in $1/D$ 
for $\tilde G \equiv (D/\theta)\tilde\Lambda_{\alpha\alpha}$ one gets
\begin{equation}\label{2.16}
\tilde G = \frac{2}{\theta}\;\frac{1}{1 + 
[ 1 + 4(h+\nu m)^2/\theta^2 + 8\tilde l_\alpha ]^{1/2} } + 
\frac{1}{D}\Delta^{(G)}
\end{equation}
where
\begin{equation}\label{2.17}
\Delta^{(G)} = \frac{2}{\theta} 
\left(
\frac{x^2}{1+x^2}g^2(x) + 
\frac{1}{4}\frac{\partial g}{\partial\tilde r} +
\frac{1}{4}\frac{\partial^2 g}{\partial\tilde r^2} 
\right)_{|\tilde r=1}
\end{equation}
is the $1/D$ correction to the Gaussian integral (\ref{2.12})
and the derivatives of $g$ are calculated with the use of (\ref{2.14}) 
and (\ref{2.15}). The \mbox{1-st} term of (\ref{2.16}) also contains 
the $1/D$ corrections due to the corresponding corrections to $m$ and 
$\tilde l_\alpha$.

Before further proceeding with the $1/D$ expansion we consider at first 
the limiting case $D\to\infty$ corresponding to the spherical model 
\cite{ber52}. In this case the quantity 
$\hat\Lambda_{\alpha\alpha}({\bf k})$ in (\ref{2.4}) can be
neglected, and one can replace in (\ref{2.13}) 
$\hat G_{\bf k}, \hat G_{\bf q} \Rightarrow G$, 
where $G=G(m_0,\tilde l_{\alpha 0})$ is given by the 
\mbox{1-st} term of (\ref{2.16}) with the zero-order quantities $m_0$ and 
\begin{equation}\label{2.18}
\tilde l_{\alpha 0} = \frac{P(G)-1}{2\theta G}; \qquad
P(G) = \,v_0\!\!\!\int\!\!\!\frac{d{\bf q}}{(2\pi)^d} \frac{1}
{1-G\lambda_{\bf q}}
\end{equation}
Here for the square lattice with the n.n. interaction the lattice integral 
$P(G)$ is given by $P(G)=(2/\pi){\bf K}(k)$ with $k\!=\!G$, ${\bf K}(k)$ 
being the elliptic integral of the \mbox{1-st} kind, and for the linear chain 
$P(G)=1/(1-G^2)^{1/2}$. 
For bipartite lattices considered here the integral $P(G)$ is the same 
for ferro- and antiferromagnets and independent of the sign of $G$. For 
this reason the sign-factor $\nu$ is dropped in the definition of 
$P(G)$ (\ref{2.18}).
Note that for 1- and 2-dimensional systems 
$P(G)$ diverges for $G\to 1$, which is the reason for the absence of the 
long-range order. Eliminating now $\tilde l_{\alpha 0}$  
and using (\ref{2.2}) in the form 
\begin{equation}\label{2.19}
\frac{m_0}{h} = \frac{G}{1-\nu G}
\end{equation}
one comes to the equation of state of the spherical model:
\begin{equation}\label{2.20}
\theta GP(G) = 1-m^2_0; 
\end{equation}
which for $h\not=0$ should be solved together with (\ref{2.19}) 
(in the general case numerically).
For low-dimensional antiferromagnets ($\nu\!=\!-1$) 
at low temperatures ($\theta\ll 1$) 
in the field region where $m^2_0<1$ the equation 
(\ref{2.20}) requires $G\cong 1$ and $P(G)\gg 1$.
For the square-lattice model for $P$ and its derivative this implies
\begin{equation}\label{4.12}
\setlength{\arraycolsep}{0.5mm}
\begin{array}{lllll}
P(G) & \cong 
& 
\displaystyle\frac{1}{\pi}\ln\left(\frac{8}{1-G}\right) & \cong 
& 
\displaystyle\frac{1-m^2_0}{\theta}          
\\
P'(G) & \cong 
& 
\displaystyle\frac{1}{\pi}\;\frac{1}{1-G} & \cong 
& 
\displaystyle\frac{1}{8\pi}\exp\left[\frac{\pi(1-m^2_0)}{\theta}\right]
\end{array}
\end{equation}
i.e. the deviation of $G$ from unity is exponentially small:
\begin{equation}\label{gexp}
G \cong 1 - 8\,\exp\left[-\frac{\pi(1-m^2_0)}{\theta}\right]
\end{equation}
For the linear chain model the corresponding result reads 
$1-G\cong \theta^2/[2(1-m^2_0)^2]$. Now with the help of (\ref{2.19})
one gets for the 
magnetization $m_0 \cong h/2$ with only exponentially 
small corrections in the 2-dimensional case due to (\ref{gexp}).
The latter is valid up to the magnetization 
saturation point $h=2$ (i.e. $H=2|J_0|$), which corresponds to the 
spin-flip field of 3-dimensional antiferromagnets. 
For the fields $h>2$ according to (\ref{2.19}) and (\ref{2.20})
$m_0\cong 1$ and $G<1$:
\begin{equation}\label{msatur}
\qquad m_0 \cong 1 - (\theta/2)GP(G); \qquad G \cong 1/(h-1)
\end{equation}
In the zero field case the antiferromagnetic susceptibility 
$\tilde\chi=\tilde\chi_\perp=\tilde\chi_z=G/(1+G)$ monotonously 
decreases with rising temperature from the value 1/2 at $T=0$ to 0 at 
$T\to\infty$, i.e. the spherical model does not describe the maximum of 
the antiferromagnetic susceptibility at $\theta\lsim 1$.

Now, returning to the $1/D$ expansion, one can express the $1/D$ 
-correction term $\Delta^{(G)}$ in (\ref{2.16}) through the variables 
of the spherical approximation:
\begin{equation}\label{2.25}
\Delta^{(G)} = 2G 
\left[ 
\frac{y-1}{2y-1} - \frac{P-1}{2(2y-1)^2} - \frac{(P-1)^2(3y-1)}{(2y-1)^3}
\right] ;
\qquad
y \equiv \frac{1}{\theta G}
\end{equation}
and represent the unknown quantities $m$, $\tilde l_\alpha$ and 
$\hat G_{\bf k}$ (see (\ref{2.13})) in the form 
\begin{equation}\label{2.21}
\setlength{\arraycolsep}{0.5mm}
\begin{array}{lll}
m & \cong & m_0 + m_1/D                
\\
\tilde l_\alpha & \cong & \tilde l_{\alpha 0} + \tilde l_{\alpha 1}/D 
\\
\hat G_{\bf k} & \cong & G + \Delta G_{\bf k}/D  
\end{array}
\end{equation}
Here the corrections $m_1$ and $\tilde l_{\alpha 1}$ can be expressed 
through  $\Delta\hat G_{\bf k}$ with the use of (\ref{2.13}) and the 
relation 
$m/h = \tilde\chi_\perp(0) = \hat G_0/(1-\nu\hat G_0)$, 
which results in
\begin{equation}\label{2.22}
\frac{m_1}{h} = 
\frac{\Delta G_0}{(1-\nu G)^2} 
\end{equation}
and 
\begin{equation}\label{2.23}
\tilde l_{\alpha 1} = 
\frac{1}{2\theta} \,v_0\!\!\!\int\!\!\!\frac{d{\bf q}}{(2\pi)^d}
\frac{\lambda^2_{\bf q} \Delta G_{\bf q}}
{(1-\nu G \lambda_{\bf q})^2}
\end{equation}
Expanding now the \mbox{1-st} term of (\ref{2.16}) 
up to the \mbox{1-st} order in 
$m_1$ and $\tilde l_{\alpha 1}$, one comes to the $1/D$ part of the 
equation (\ref{2.4}) in the dimensionless form 
\begin{equation}\label{2.24}
\Delta G_{\bf k} + \frac{m_0}{\nu h} \;
\frac{2m_0^2y}{2y-1}\Delta G_0 + \frac{G^2}{2y-1} 
\,v_0\!\!\!\int\!\!\!\frac{d{\bf q}}{(2\pi)^d}\frac{\lambda^2_{\bf q}\Delta 
G_{\bf q}}
{(1-\nu G \lambda_{\bf q})^2} =
\Delta^{(G)} + \Delta^{(1/D)}_{\bf k} ,
\end{equation}
where the quantity $\Delta^{(1/D)}_{\bf k}$ is 
the nonvanishing in the limit $D\to\infty$ part of \linebreak
\mbox{$D\cdot(D/\theta)\hat\Lambda^{(1/D)}_{\alpha\alpha}({\bf k})$} 
(see Appendix).
The solution of the integral equation (\ref{2.24}) has the form
\begin{equation}\label{2.26}
\Delta G_{\bf k} = \Delta G_0 + M_{\bf k}
\end{equation}
where
\begin{equation}\label{2.27}
M_{\bf k} = \Delta^{(1/D)}_{\bf k} - 
\Delta^{(1/D)}_0
\end{equation}
and $\Delta G_0$ is given by
\begin{eqnarray}\label{dg0}
\Delta G_0  = 
\left\{
(2y-1)(\Delta^{(G)} + \Delta^{(1/D)}_0) -
G^2 \,v_0\!\!\!\int\!\!\!\frac{d{\bf q}}{(2\pi)^d}\frac{\lambda^2_{\bf 
q}M_{\bf q}}
{(1-\nu G \lambda_{\bf q})^2} 
\right\}\cdot
\\
\left\{
GP'\!+\! P(G)\! + \frac{2m^2_0y}{1-\nu G}
\right\}^{-1} \nonumber
\end{eqnarray}
where $P'\equiv dP(G)/dG$.
Now, calculating the quantity $\Delta^{(1/D)}_{\bf k}$
(see Appendix) and introducing the function
\begin{equation}\label{3.9}
r_{\bf q} = \,v_0\!\!\!\int\!\!\!\frac{d{\bf p}}{(2\pi)^d} 
g_{\bf p} g_{\bf p-q} ; \qquad
g_{\bf p} \equiv \frac{1}{1-\nu G \lambda_{\bf p}}
\end{equation}
one arrives after numerous canscellations at the final results
for $\Delta G_0$ and $M_{\bf k}$:
\begin{eqnarray}\label{3.11}
\Delta G_0 = 2G
\left\{
1 - \left[ GP'\!+\!P\!+\!2m^2_0y) \right]
v_0\!\!\!\int\!\!\!\frac{d{\bf q}}{(2\pi)^d} \frac{g_{\bf q}}
{\tilde r_{\bf q}} +
m^2_0y 
\,v_0\!\!\!\int\!\!\!\frac{d{\bf q}}{(2\pi)^d} \frac{g^2_{\bf q}}
{\tilde r_{\bf q}} 
\right. 
\\[-1ex]
\left.
+\,\frac{G}{2}
\,v_0\!\!\!\int\!\!\!\frac{d{\bf q}}{(2\pi)^d} 
\frac{r'_{\bf q}}{\tilde r_{\bf q}} 
\right\}
\left\{
GP'\!+\! P\! + \frac{2m^2_0y}{1-\nu G}
\right\}^{-1}                  
\nonumber
\end{eqnarray}
and 
\begin{equation}\label{3.12}
M_{\bf k} = 
2G\,v_0\!\!\!\int\!\!\!\frac{d{\bf q}}{(2\pi)^d}
\frac{g_{\bf q}-g_{\bf q-k}}
{\tilde r_{\bf q}}
\end{equation}
where $\tilde r_{\bf q} \equiv r_{\bf q} + 2m^2_0yg_{\bf q}$ and 
$r'_{\bf q} \equiv \partial r_{\bf q}/\partial G$. The similar results 
obtained earlier \cite{abe7273,abe7377} by another method for the 
particular ferromagnetic case were used for 
the investigation of the 
phase transition in 3-dimensional ferromagnets. It is interesting to note 
that the function $r_{\bf q}$ (\ref{3.9}) 
is (like $P(G)$) identical for ferro- and antiferromagnets and has a 
singularity at $G \!\to\! 1$ and ${\bf q} \!\to\! 0$. In 
contrast, the quantity $\tilde r_{\bf q}$ entering (\ref{3.11}) 
and (\ref{3.12}) has for antiferromagnets one more singularity at 
$G \!\to\! 1$ and ${\bf q} \!\to\! {\bf b}$ (${\bf b}$ is the inverse lattice 
vector) due to $g_{\bf q}$, which disappears, however, 
in zero magnetic field ($m_0 = 0$). This is a formal mechanism 
responsible for the 
singular behavior of the susceptibility $\chi(H,T)$ of the 
low-dimensional antiferromagnets in the limit $H,T \to 0$ discussed in 
the Introduction.

Before proceeding with the application of the results obtained to the 
concrete systems it is worth to note some general properties of the 
{\bf k}-dependent spin-spin correlation function (see (\ref{2.1}) and 
(\ref{2.13})) that can 
essentially simplify the consideration in the low-temperature range. 
In particular, for 2-dimensional ferromagnets in the spherical limit 
the quantity $G$ (\ref{gexp}) is exponentially close to 
unity at $h=0$ and low temperatures, which implies exponentially small 
gap in the spin-wave spectrum. Since this property cannot be changed 
with taking into account $1/D$ corrections, the quantity $\Delta G_0$ in 
(\ref{2.26}) should be also exponentially small. This is 
physically clear and can be confirmed by the direct analysis 
\cite{garst94} of the results obtained. On the other hand, the 
{\bf k}-dependent contribution $M_{\bf k}$ in (\ref{2.26}) has not to be 
exponentially small at low temperatures and ${\bf k}\!\not=\!0$. In fact, 
the value $M_{\bf b}$ determines the $1/D$ correction to the {\it 
antiferromagnetic} susceptibility in zero magnetic field 
\cite{garst94}, which can be expanded in powers of $\theta \ll 1$. 
Thus, by calculation of such quantities of 2-dimensional magnetic 
systems at low temperatures, which are not exponentially small, one can 
use only the quantity $M_{\bf k}$ (\ref{3.12}), being much simpler 
than the expression for $\Delta G_0$ (\ref{3.11}). This means that only 
{\bf k}-dependent diagrams for the compact part of the spin-spin 
correlation function $\hat\Lambda_{\alpha\alpha}({\bf k})$ should be 
taken into account in the low-temperature range, which is a clear 
advantage of the diagrammatic $1/D$ expansion in comparison with the 
earlier version \cite{abe7273,abe7377}. The considerations above can be 
extended also on 2-dimensional {\it antiferromagnets} at low temperatures in 
the field region $h<2$ ($H< 2|J_0|$), where the magnon gap is 
exponentially small. Here the quantity $\hat G_{\bf b}$ in 
(\ref{2.13}) should be exponentially close to unity in all orders in 
$1/D$. Consequently, the quantity $\Delta G_0$ contributing to the 
magnetization and susceptibility of an antiferromagnet (see (\ref{2.22})) 
is given according to (\ref{2.26}) by 
$\Delta G_0 =  \Delta G_{\bf b} - M_{\bf b} \cong -M_{\bf b}$.
In the next sections we apply the results of the \mbox{1-st} order in 
$1/D$ obtained above 
to the analysis of the equation of state $m(H,T)$ of 1- and 
2-dimensional classical antiferromagnets.

\section{The linear chain classical spin model} 
\eqreset

For the linear chain model $\lambda_k \equiv J_k/J_0 = \cos(a_0k)$, 
and the integrals 
(\ref{3.9}) and (\ref{3.11}) can be calculated analytically. One gets 
$r_q = 2P(G)/(2-G^2-G^2\lambda_q)$ with $P(G) = 1/(1-G^2)^{1/2}$ 
and 
\begin{eqnarray}\label{4.1}
\Delta G_0 = \frac{2G(1-m^2_0)(1-G^2)^{3/2}}{1+m^2_0+2m^2_0\nu G}
\left\{ 
1 + \frac{3}{2}P(G) -
\frac{5+3m^2_0-2m^2_0G^2}{2[1+m^2_0(1-G^2)]}F  
\right.                                              \\
\left.
+\,\frac{1}{2}\;\frac{1-F}{1-m^2_0(1-\nu G)} 
\left[ 
\frac{1-m^2_0}{1-G^2}(1-\nu G) - 3(1-m^2_0) - 2\nu Gm^2_0 
\right]
\right\}                                               \nonumber
\end{eqnarray}
where 
\begin{equation}\label{4.2}
F = \frac{1+m^2_0(1-G^2)}
{\left[
(1+m^2_0)^2(1-G^2)+2G^2m^2_0(1-m^2_0)(1-\nu G) 
\right]   ^{1/2}}
\end{equation}
The magnetization $m_0$ and the parameter $G$ of the spherical model 
in (\ref{4.1}) and (\ref{4.2}) are given by the solution of  
(\ref{2.19}) and (\ref{2.20}). It can be shown that 
in zero magnetic field the results 
for the susceptibility $\chi$ are equivalent to those obtained by the 
expansion of the exact solution \cite{sta69} up to the \mbox{1-st} order in $1/D$. 
In the low field and temperature limit $h,\theta\ll 1$ one has $m_0 
\cong h/2 \ll 1$, $ G \cong 1-\theta^2$ and hence 
$F \cong (h^2+\theta^2)^{-1/2} \gg 1$. Taking into acount the leading 
contribution into $\Delta G_0$ given by the \mbox{1-st} term in square 
brackets, one gets 
with the use of (\ref{2.21}) and (\ref{2.22}) the following result
\begin{equation}\label{4.3}
\tilde\chi_\perp = \frac{m}{h} = \frac{1}{2}
\left[
1 + \frac{1}{D}
\left(
-\frac{\theta}{(h^2+\theta^2)^{1/2}} + \theta + O(\theta^2)
\right)
\right]
\end{equation}
It can be seen that for $h=0$ the susceptibility $\tilde\chi_\perp$ 
decreases with lowering temperature due to the term $\theta$ in 
(\ref{4.3}) and attains the value $\tilde\chi_\perp=(1/2)(1-1/D)$ at 
$\theta=0$. If $h\not=0$, then at $\theta=h^{2/3}$ the value of 
$\tilde\chi_\perp$ attains a minimum and then rises to 1/2 at 
$\theta=0$. Note that the singular term in (\ref{4.3}) becomes
of order unity at $\theta \sim h \ll h^{2/3}$, which is one more
characteristic temperature.
Such a qualitative behavior of the susceptibility of a 
low-dimensional classical antiferromagnet is in accord with the physical 
considerations made in the Introduction. 
The longitudinal susceptibility $\tilde\chi_z$ 
calculated with the help of (\ref{2.3}) and (\ref{4.3}) has the form
\begin{equation}\label{4.4}
\tilde\chi_z = \frac{\partial m}{\partial h} = \frac{1}{2}
\left[
1 + \frac{1}{D}
\left(
-\frac{\theta^3}{(h^2+\theta^2)^{3/2}} + \theta + O(\theta^2)
\right)
\right]
\end{equation}
This expression has the minimum at 
$\theta \cong 3^{1/3}h^{2/3}\gg h$, 
and also the maximum at $\theta \cong 3^{-1/2}h^{3/2} \ll h$ 
(the 3-rd characteristic temperature)
where 
$\tilde\chi_z \cong 1/2 + (2/D)(h/3)^{3/2} > 1/2$. The 
susceptibilities $\tilde\chi_\perp(h,\theta)$ and 
$\tilde\chi_z(h,\theta)$ are represented as functions of temperature for 
some field values in figs.3,4. 
\begin{figure}[p]
\unitlength1cm
\begin{picture}(15,8)
\centerline{\epsfig{file=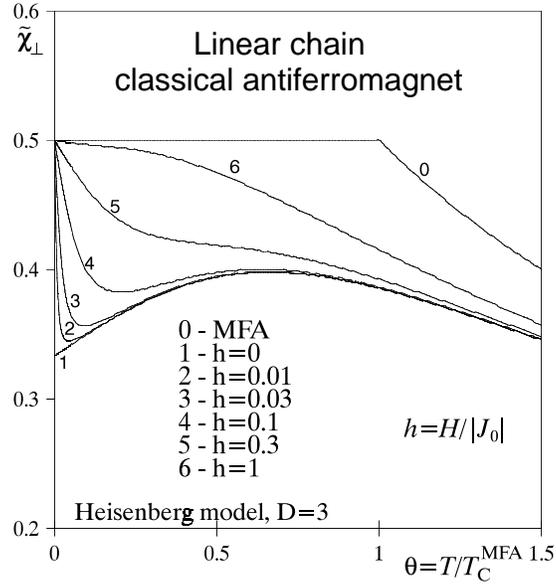,angle=0,width=9cm}}
\end{picture}
\caption{  
Temperature dependences of the transverse susceptibility 
$\tilde\chi_\perp=m/h$
of the l.c. Heisenberg antiferromagnet 
for different magnetic fields in the \mbox{1-st} order in $1/D$.
}
\end{figure}
\begin{figure}[p]
\unitlength1cm
\begin{picture}(15,8)
\centerline{\epsfig{file=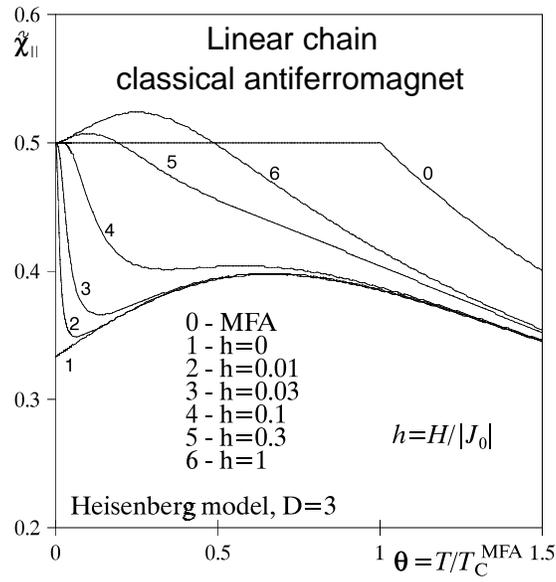,angle=0,width=9cm}}
\end{picture}
\caption{  
Temperature dependences of the longitudinal susceptibility 
$\tilde\chi_z=\partial m/\partial h$.
}
\end{figure}\par
An interesting feature of the susceptibility $\tilde\chi_z$ 
manifests itself in the $1/D$ approximation in the low temperature limit 
($\theta \ll 1$) in the vicinity of the magnetization saturation point 
$h=2$ (i.e. $H=H_c\equiv 2|J_0|$). In the spherical limit, adopting 
$m_0=1-\delta m_0$ and $G=1-\delta G$ with $\delta m_0, 
\delta G \ll 1$, 
one can simplify the equations (\ref{2.19}) and (\ref{2.20}) for the 
linear chain antiferromagnet to $\theta/(2\delta G)^{1/2}=2\delta m_0$ and 
$2\delta m_0-\delta G = 2-h$, which results in the following 
equation for $\delta m_0$ in the scaled form
\begin{equation}\label{4.5}
x - 1/(16x^2) = x_0; \qquad 
x \equiv \delta m_0/\theta^{2/3}; \qquad
x_0 \equiv (2-h)/(2\theta^{2/3})
\end{equation}
This equation describes the temperature-induced rounding of the 
transition between the dependences $m_0\cong h/2$ and $m_0\cong 1$ in 
the small field interval $|2-h| \sim \theta^{2/3}$.
Now, the $1/D$ correction $m_1$ determined for $|2-h|,\theta \ll 1$ 
from (\ref{2.22}) and (\ref{4.1}), (\ref{4.2}) has the form
\begin{equation}\label{4.7}
m_1 \cong \delta m_0 (3 - 3Y^{-1/2} + Y^{1/2})/Y ; \qquad
Y \equiv 1 + \delta m_0/\delta G = 1 + 8x^3
\end{equation}
where $x$ is the solution of (\ref{4.5}). In the limiting cases one gets 
from (\ref{4.6}) and (\ref{4.7}) for the 
magnetization $m=m_0+m_1/D$ the following results
\begin{equation}\label{4.6}
\renewcommand{\arraystretch}{1.5}
m \cong 
\left\{
\begin{array}{ll}
\displaystyle \frac{h}{2} + \frac{1}{D}\,\frac{\theta}{2(2-h)^{1/2}} -
\frac{\theta^2}{[2(2-h)]^2} ;              & \theta^{2/3} \ll 2-h \ll 1 
\\[1.5ex]
\displaystyle 1 - \left( \frac{\theta}{4} \right)^{2/3} \!
\left[
1 - \frac{1}{D}\left( \frac{2}{3} \right)^{1/2}\!\!(6^{1/2}\!-\!1)
\right]                                  ;
                                           &     h=2 
\\
\displaystyle 1-\frac{\theta}{[8(h-2)]^{1/2}} 
\left( 1 - \frac{1}{D}\right) ;            &     \theta^{2/3} \ll h-2 \ll 1
\end{array}
\right.
\end{equation}
It can be seen from (\ref{4.6}) that in the field region 
below the saturation point $h=2$ the 
temperature-dependent correction to $m$ is positive.
Accordingly, the susceptibility $\tilde\chi_z=\partial m/\partial h$ is 
greater than 1/2 in this region, but the effect is not great. With the 
use of (\ref{4.7}) one can 
show that for $D\!=\!3\; \tilde\chi_{z,max}\!=\!0.518$ at 
$2-h=2.80\cdot\theta^{2/3}$.
The field dependences of the normalized susceptibility 
$\tilde\chi_z$ of the 1-dimensional classical antiferromagnet are 
represented for different temperatures in fig.5. It is
interesting to note that a qualitatively similar field
dependence of the susceptibility with a logarithmic singularity 
at small fields was found in ref. \cite{gri64} for the 
{\it quantum} linear chain Heisenberg
antiferromagnet with $S=1/2$ at $T=0$. There are no physical
comments to this effect in ref. \cite{gri64}, but it seems now
rather clear that the origin of this low-field singularity of
a quantum antiferromagnet is also the orientation of sublattices
perpendicular to the field, the quantum effects playing here the role
of some "residual temperature".  
\begin{figure}
\unitlength1cm
\begin{picture}(15,8)
\centerline{\epsfig{file=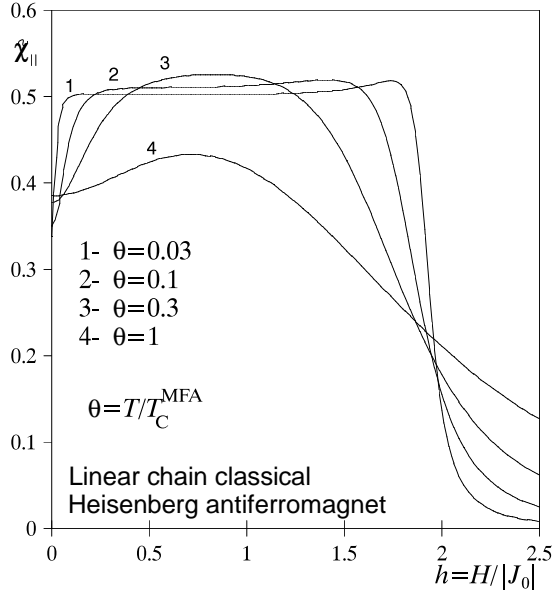,angle=0,width=9cm}}
\end{picture}
\caption{  
Field dependences of the longitudinal susceptibility of the l.c. 
classical Heisenberg antiferromagnet for different temperatures.
}
\end{figure}

\section{The square-lattice classical antiferromagnet}
\eqreset

For 2-dimensional lattice the integrals in $\Delta G_0$ (\ref{3.11}) 
cannot be calculated analytically. For the convenience of the analysis 
at low temperatures and numerical calculations we introduce instead of 
the strongly singular $r_{\bf q}$ (\ref{3.9}) 
the weak-singular function \cite{garst94}
\begin{equation}\label{4.9}
\psi_{\bf q} \equiv \frac{1}{G}
\,v_0\!\!\!\int\!\!\!\frac{d{\bf p}}{(2\pi)^d}
\frac{\lambda_{\bf q} - G\lambda_{\bf p} 
\lambda_{\bf p-q}}
{(1-G\lambda_{\bf p})(1-G\lambda_{\bf p-q})} =
\frac{1}{G^2}
\left[
2P(G) -1 - (1-G\lambda_{\bf q})r_{\bf q}
\right]
\end{equation}
in which the divergences of the integrand at ${\bf p}=0$ and
${\bf p=q}$ at $G \cong 1$ are partially compensated by the
nullification of the numerator. In the longwavelength region 
the function $\psi_{\bf q}$ has the form \cite {garst94}:
\begin{equation}\label{psi}
\renewcommand{\arraystretch}{2}
\psi_{\bf q} \cong 
\left\{
\begin{array}{ll}
\displaystyle
\frac{2}{\pi}\ln\frac{8}{1-G} - 1 - \frac{1}{\pi} ;
& x \equiv (a_0 q)^2 \ll 1-G
\\
\displaystyle
\frac{2}{\pi}\ln\frac{8}{x} ;
& 1-G \ll x \ll 1
\end{array}
\right.
\end{equation}
and its derivative $\psi'_{\bf q} \equiv \partial\psi_{\bf q}/\partial G$
is given by
\begin{equation}\label{dpsi}
\renewcommand{\arraystretch}{2}
\psi'_{\bf q} \cong 
\left\{
\begin{array}{ll}
\displaystyle
\frac{2}{\pi}\,\frac{1}{1-G} ;
& x \ll 1-G
\\
\displaystyle
-\frac{2}{\pi x}\ln\frac{x}{1-G} ;
& 1-G \ll x \ll 1
\end{array}
\right.
\end{equation}
At low temperatures at the corners of the Brillouin zone 
$\psi_{\bf q} = -1 + O(1-G)$.
In terms of $\psi_{\bf q}$ the function $\Delta G_0$ (\ref{3.11}) can be  
written as
\begin{eqnarray}\label{4.10}
\Delta G_0 = 2G
\left\{ 
(1\!-\!\nu)\frac{GP'\!+\!P\!+\!3m^2_0y}{2y-1}
\,v_0\!\!\!\int\!\!\!\frac{d{\bf q}}{(2\pi)^d}
\frac{1}{1-a\psi_{\bf q}}
\frac{G\lambda_{\bf q}}{1+\bar G_{\bf q}\lambda_{\bf q}} \nonumber
\right.
\\
\left.
+\,\frac{G}{2}
\,v_0\!\!\!\int\!\!\!\frac{d{\bf q}}{(2\pi)^d}
\frac{a\psi'_{\bf q}}{1-a\psi_{\bf q}}
\frac{1+G\lambda_{\bf q}}{1+\bar G_{\bf q}\lambda_{\bf q}} 
- \frac{3y-1}{2y-1}
\,v_0\!\!\!\int\!\!\!\frac{d{\bf q}}{(2\pi)^d}
\frac{1}{1-a\psi_{\bf q}}
\frac{1+G\lambda_{\bf q}}{1+\bar G_{\bf q}\lambda_{\bf q}}  
\right.                          
\\
\left.
+\,\frac{3}{2} + P(G)
- \frac{1}{2}
\,v_0\!\!\!\int\!\!\!\frac{d{\bf q}}{(2\pi)^d}
\frac{1}{1+\bar G_{\bf q}\lambda_{\bf q}}  
\right\}  
\left\{
GP'\!+\! P(G)\! + \frac{2m^2_0y}{1-\nu G}
\right\}^{-1}    \nonumber
\end{eqnarray}
where $y\equiv 1/(\theta G)$, $a \equiv G^2/(2y-1)$ and 
\begin{equation}\label{4.11}
\bar G_{\bf q} \equiv G 
\left[
1 - \frac{2(1-\nu)m^2_0y}{(2y-1)(1-a\psi_{\bf q})}
\right]
\end{equation}
For the quantity $M_{\bf k}$ (\ref{3.12}) one gets in a similar way
\begin{equation}\label{mkpsi}
M_{\bf k} = \frac{2G}{2y-1}
\,v_0\!\!\!\int\!\!\!\frac{d{\bf q}}{(2\pi)^d}
\frac{ ( 1-G^2\lambda^2_{\bf q} )( g_{\bf q}-g_{\bf q-k} ) }
{ ( 1-a\psi_{\bf q} )( 1+\bar G_{\bf q}\lambda_{\bf q} ) }
\end{equation}
Putting ${\bf k=b}$ in (\ref{mkpsi}) in the antiferromagnetic case
($\nu\!=\!-1$) and 
taking into account only the exponentially great terms with $P'(G)$ 
in $\Delta G_0$ (\ref{4.10}) at low 
temperatures in the field range $h<2$,  one arrives at the result 
\begin{equation}\label{deltag0mb1}
\Delta G_0 \cong -M_{\bf b} = 4a
\,v_0\!\!\!\int\!\!\!\frac{d{\bf q}}{(2\pi)^d}
\frac{1}{1-a\psi_{\bf q}}
\frac{\lambda_{\bf q}}{1+\bar G_{\bf q}\lambda_{\bf q}}
\end{equation}
which confirms the conjecture made at the end of Section 2. 
Since at $\theta \ll 1$ the quantity 
$a \cong (\theta/2)/(1-\theta/2) \ll 1$ and due to (\ref{psi})
the functions $\psi^n_{\bf q}$ are integrable ones, one can
expand (\ref{deltag0mb1}) in powers of $a\psi_{\bf q}$ and then
of $\theta$ to get the development of the $1/D$ correction to
the magnetization $m_1$ (\ref{2.22}) at
low temperatures. In the lowest order in $\theta$ one gets
\begin{equation}\label{4.13}
m_1 = \frac{h\theta}{2}\,\frac{1-P(\bar G)}{1-2m^2_0}
\end{equation}
where $\bar G \cong G(1-2m^2_0)$ and $m_0 \cong h/2$. 
This correction is negative for $h < 2^{1/2}$ and positive for 
$2^{1/2} < h \lsim 2$.
In the case $h\not=0$ the quantity $1-\bar G$ can be interpreted
as proportional to the field-induced gap of the out-of-plane
spin waves, which makes the lattice integral $P(\bar G)$
in (\ref{4.13}) not divergent at low temperatures. The more
detailed physical interpretation of the spin wave dynamics in
low-dimensional magnets requires, however, the dynamical
generalization of the diagram technique used here. 
In the small field region, where $\bar G \cong 1$, with the help
of (\ref{2.21}) and (\ref{2.2}) 
in the \mbox{1-st} order in $1/D$ one gets
\begin{equation}\label{4.14}
\tilde\chi_\perp = \frac{m}{h} = \frac{1}{2}
\left[
1 + \frac{1}{D}
\left(
-\frac{\theta}{\pi}\ln\left(\frac{8}{1-G+h^2/2}\right)
 + \theta + O(\theta^2)
\right)
\right]
\end{equation}
where $G$ is given by (\ref{gexp}). 
In the case $h=0$ the ln-term in (\ref{4.14}) is identically equal to $-1$, 
and $\tilde\chi_\perp \to (1/2)(1-1/D)$ in the 
limit $\theta \to 0$. For a whatever small field $h\not=0$ this term goes to 
zero with $\theta \to 0$, and $\tilde\chi_\perp \to 1/2$. 
The transition to the regime, where  
the magnetic field exerts the influence on the susceptibility of a 
2-dimensional antiferromagnet, is sharp due to the strong exponential
temperature dependence of $G$ (\ref{gexp}) and occures at the temperature
\begin{equation}\label{thetstar}
\theta \cong \theta^* = \frac{\pi}{2\,\ln(4/h)}
\end{equation}
Note that the value of $\theta^*$ is for $h \ll 1$ much 
larger than the corresponding characteristic temperatures 
in the 1-dimensional case. The 
longitudinal susceptibility of the s.q. classical antiferromagnet has 
the form
\begin{equation}\label{4.15}
\tilde\chi_z = \frac{1}{2}
\left[
1 + \frac{1}{D}
\left(
-\frac{\theta}{\pi}\ln\left(\frac{8}{1-G+h^2/2}\right) +
\frac{\theta}{\pi}\frac{h^2}{1-G+h^2/2} + \theta + O(\theta^2)
\right)
\right]
\end{equation}
where the addditional in comparison to (\ref{4.14}) term is
not very essential at low fields in contrast to the
1-dimensional case (see (\ref{4.4}). 
The temperature dependences of $\tilde\chi_\perp$ and $\tilde\chi_z$ in 
the magnetic field obtained by the numerical solution of the equations 
(\ref{2.19}) and (\ref{2.20}) and the numerical integration in 
(\ref{4.10})  are represented in figs.6 and 7.
\begin{figure}[p]
\unitlength1cm
\begin{picture}(15,8)
\centerline{\epsfig{file=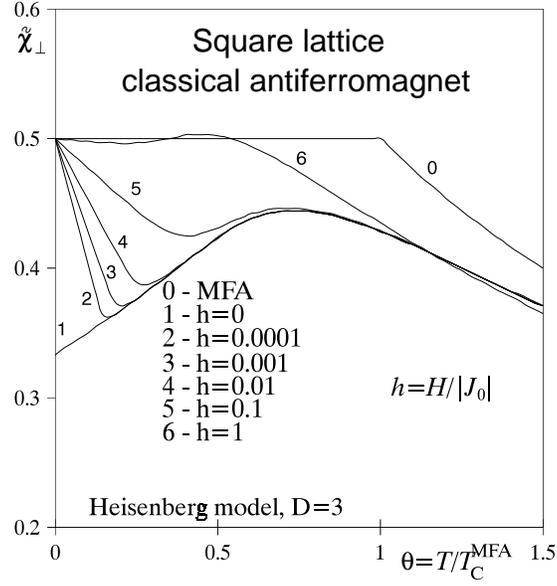,angle=0,width=9cm}}
\end{picture}
\caption{  
Temperature dependences of the transverse susceptibility of the s.l. 
classical Heisenberg antiferromagnet for different magnetic fields
in the \mbox{1-st} order in $1/D$. 
}
\end{figure}
\begin{figure}[p]
\unitlength1cm
\begin{picture}(15,8)
\centerline{\epsfig{file=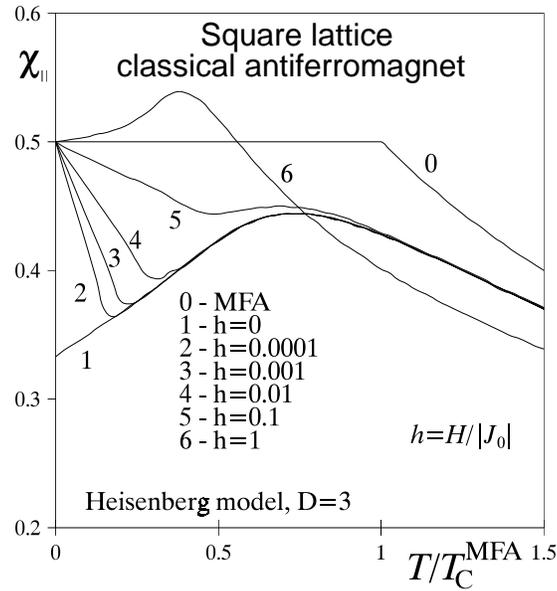,angle=0,width=9cm}}
\end{picture}
\caption{  
Temperature dependences of the longitudinal susceptibility. 
}
\end{figure}\par
The $1/D$ correction to the magnetization (\ref{4.13}) increases
with approaching the magnetization saturation point $h=2$ since
here $\bar G \cong -1$ and $P(\bar G) \gg 1$. But the formula (\ref{4.13})
becomes insufficient in this region, because in (\ref{4.10})
the integral with  $\psi'_{\bf q}$ (see (\ref{dpsi}))
becomes for $\bar G \cong -1$  comparable with the one with
$P'(G)$ due to the great longwavelength contribution, 
and one should use the complicated analytical expression for
$\psi'_{\bf q}$ for $x \sim 1\!-\!G$ \cite {garst94}. In
contrast, the quantity $a\psi_{\bf q}$ in the denominators in (\ref{4.10})
and (\ref{4.11}) can be neglected in the whole field region,
since in the low-temperature range 
$a\psi_0 \cong \theta P(G) \cong 1\!-\!m^2_0$.  
In the field region above the saturation point ($h>2$) at low 
temperatures
$m_0 \cong 1$ and in (\ref{4.11}) $\bar G_{\bf q} \cong -G$.
Neglecting the terms $\theta P' \ll 1$ in 
$\Delta G_0$ (\ref{4.10}), 
with the use of (\ref{msatur}) for the total magnetization 
$m=m_0+m_1/D$ one gets 
\begin{equation}\label{msw}
m \cong 1 - \frac{\theta}{2}\left(1-\frac{1}{D}\right)GP(G);
\qquad G \cong \frac{1}{h-1}
\qquad (h>2, \theta \ll 1)
\end{equation}
This result is the exact expression for the leading correction to the 
magnetization of a classical antiferromagnet in the spin-flip phase 
($H>2|J_0|$) in the low-temperature limit, which can be obtained 
independently with the help of the lowest-order spin wave theory. In the 
framework of the diagram technique for classical spin systems used here 
this corresponds to taking into account only the simplest diagram for 
the magnetization 
$m$ with one integration over the Brillouin zone (i.e. the one 
analogous to the 2-nd diagram in fig.1,b). The derivation of the formula 
(\ref{msw}) is trivial, because the ground state of the system has no 
spontaneous symmetry breaking and the magnon spectrum has a gap. 
However, with the approach to $h=2$ in (\ref{msw}) $G \to 1$ and for 
low-dimensional systems the temperature correction to $m$ diverges. In 
the region $h < 2$ the situation becomes complicated, and to obtain 
finite results for the thermodynamic quantities one has to take into account 
some infinite series of diagrams, which is exemplified by the $1/D$ 
expansion described above. With the help of (\ref{4.13}) and
(\ref{msw}) one can write down the expressions for the
magnetization of the 2-dimensional antiferromagnet on both sides
of the magnetization saturation point $h=2$ excluding a small
intermediate region:   
\begin{equation}\label{msq}
\renewcommand{\arraystretch}{2}
m \cong 
\left\{
\begin{array}{ll}
\displaystyle 
\frac{h}{2} + \frac{\theta}{\pi D}\ln\frac{4}{2-h};
& 
\displaystyle
\theta \ln\frac{1}{\theta} \ll 2-h \ll 1
\\
\displaystyle
1 - \frac{\theta}{2\pi}\left(1-\frac{1}{D}\right)\ln\frac{8}{h-2};
&
\displaystyle
\theta \ln\frac{1}{\theta} \ll h-2 \ll 1
\end{array}
\right.
\end{equation}
The latter results are analogous to those for the linear chain model 
(\ref{4.6}) that could be also obtained in the same way like here. 
The normalized susceptibility 
$\tilde\chi_z$ of the 2-dimensional antiferromagnet is greater then 1/2 
below the saturation point $h=2$, too, and the maximal value of 
$\tilde\chi_z$ is greater than that for the linear chain (see fig.8). 
The latter can be explained by the fact that for a square lattice there 
is no competing {\it negative} contribution to $m$ of the zeroth 
order in $1/D$, as is the case for the linear chain (see (\ref{4.6})).
\begin{figure}
\unitlength1cm
\begin{picture}(15,8)
\centerline{\epsfig{file=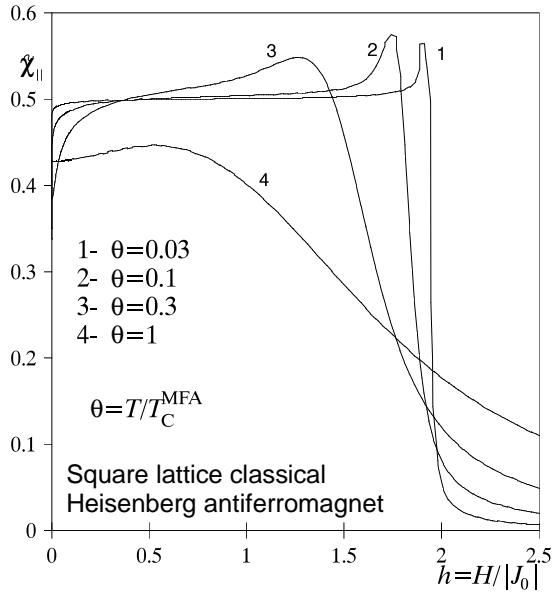,angle=0,width=9cm}}
\end{picture}
\caption{  
Field dependences of the longitudinal susceptibility of the s.l. 
classical Heisenberg antiferromagnet for different temperatures. 
}
\end{figure}

\section{Discussion}
\eqreset

In this article the $1/D$ expansion of the physical 
quantities of low-dimensional classical $D$-vector models in the whole 
range of temperatures and magnetic fields was developed, 
the results obtained being 
valid for both ferro- and antiferromagnets. For the calculation of the 
susceptibility and the field-induced magnetization of ferromagnets at 
low temperatures the method is, however, not very efficient, because these 
quantities are singular at $T\to 0$. In ref. \cite {garst94} was shown 
that at low temperatures the $1/D$ correction to the 
susceptibility of a 2-dimensional ferromagnet becomes greater than its value 
in the zeroth order in $1/D$, which means that $D$ enters the argument 
of the exponentially great expression for $\chi \propto 1/(1-\hat G_0)$ 
(comp. (\ref{4.12})). This is in accord with the results of the 
RG-approach of ref. \cite {fukoya83}, but does not allow to obtain 
accurate results in the framework of the $1/D$ expansion. 

On the other hand, the $1/D$ expansion proves to be a very good 
instrument for the description of {\it noncritical} characteristics 
of low-dimensional magnets, such as the magnetization and susceptibility 
of antiferromagnets and the energy and heat capacity of both ferro- and 
antiferromagnets. For the latter the zero-field results (identical in 
both cases) were obtained in ref. \cite {garst94}, and their 
generalization to the case with the magnetic field with the use of the 
methods developed here makes no difficulties. The most remarcable 
feature of the results obtained with the help of the $1/D$ expansion is 
that they describe the maximum in the temperature dependence of the 
zero-field antiferromagnetic susceptibility and its singular behavior at 
$H,T \to 0$. The former is the result of taking into account the 
diagrams with {\it double} integrations over the Brillouin zone, which 
was not made in any of the preceding theories. This means allowing 
for the wavevector dependence of the compact part of the spin-spin 
correlation function (\ref{2.1}) as well, or, in the other words,
going 
beyond the Ornstein-Zernike form for $\chi_{\bf k}$. It is worth noting, 
however, that at low temperatures ($\theta \ll 1$) the results simplify 
to some expression with only {\it one} integration over the Brillouin 
zone (see (\ref{4.13})) which corresponds to taking into account
the out-of-plane magnons having the field-dependent gap. 

An intriguing property of the $1/D$ expansion is that it leads to the 
exact results for the {\it noncritical} characteristics of low-dimensional 
magnets at low temperatures. In particular, the physically 
expected limiting value 
of the antiferromagnetic susceptibility at $T\to 0$ and $H=0$ is exactly 
recovered and the singularity of $\chi(H,T)$ at $H,T \to 0$ is entirely 
explained in the \mbox{1-st} order in $1/D$. The latter it is unlikely to 
be essentially changed in the next orders in $1/D$. All the examples 
considered up to now suggest that the coefficients in the expansions of 
the noncritical quantities in powers of $\theta$ are polinoms in $1/D$ 
(see ref. \cite {garst94}). If it is true, then the up to now not 
available low-temperature expansions of these 
quantities can be obtained with the help of the $1/D$ 
expansion! Further it should imply that there is some method of 
derivation of these low-temperature expansions without using the 
$1/D$ expansion. The search for such a method is planned for the nearest 
future.

It would be very interesting to compare the results of the $1/D$ 
expansion with results obtained by other methods. In particular, 
for the energy of a square-lattice classical Heisenberg magnet the MC 
simulations were made by Shenker and Tobochnik \cite {she80} (see the 
comparison in ref. \cite {garst94}), but the antiferromagnetic 
susceptibility was simulated by various researches 
only for a quantum model with $S=1/2$. 
As concerns the 2-dimensional model with $D=2$, the $1/D$ expansion 
cannot, naturally, describe the Kosterlitz--Thouless transition, which 
occurs in this system. But one can expect that the general features of 
the temperature dependence of the antiferromagnetic susceptibility in 
magnetic fields described by the $1/D$ expansion are inherent in this 
model, too. Moreover, in the magnetic field the behavior of the 
antiferromagnetic model with $D=2$ should simplify, because the magnetic 
field lifts the spontaneous symmetry breaking and induces the gap of   
spin fluctuations.
In this case at low temperatures it is enough to take 
into account only the lowest order diagram of the spin-wave theory that 
is naturally contained in the $1/D$ expansion in the \mbox{1-st} order in $1/D$ 
(comp. (\ref{msw})). 
It should be stressed that for the model with $D=2$ the effects in the
temperature and field dependence of the antiferromagnetic
susceptibility discussed in this paper should be the most
strongly revealed.
For the Heisenberg ($D=3$) antiferromagnet the 
behavior in the magnetic field can be 
more complicated, than for $D=2$. As we have seen, at low temperatures 
even small magnetic field forces the spins to lie perpendicular to it. 
This decreases the effective number of spin components from $D=3$ to 
$D=2$ and should lead to the Kosterlitz--Thouless transition with 
disappearance of the gap. But it should not change essentially the results  
for the susceptibility in this region, since in the expression for the 
$1/D$ correction to the magnetization (\ref{4.13}) enters the  
gap of the out-of-plane spin waves, which cannot disappear due
to the Kosterlitz--Thouless transition.

It should be also mentioned that the $D$-component vector model 
\cite {sta6871}
considered in this article can be generalized for hamiltonians with 
the spin anisotropy. For example, one can consider the so-called $n\!-\!D$ 
model \cite {garlut8486}, in which only $n$ from the total $D$ spin 
components are coupled by the exchange interaction. In this sence the 
$x\!-\!y$ model ($D\!=\!3,\, n\!=\!2$) is something different from 
the plane rotator model ($D\!=\!n\!=\!2$). 
It should be noted that the quantitis $n$ and $D$ play the
different roles: the well-known expansion of the critical indices of 
3-dimensional systems is the expansion in $1/n$, and the expansion 
developed here for low-dimensional systems is the $1/D$ expansion. The 
results of the present article can be generalized for the $n\!-\!D$ model, 
as well as for more general models with anisotropic spin 
interactions.

\section*{Acknowledgments}

The author thanks Hartwig Schmidt for valuable discussions. 
The financial support of Deutsche Forschungsgemeinschaft unter contract Schm
397/5-1 is greatfully acknowledged.

\vspace{1cm}
{\LARGE Appendix}

\renewcommand{\theequation}{A.\arabic{equation}}

\section*{The \mbox{\boldmath 1/D} diagrams}
\eqreset

The additional diagrams constituting 
$\hat\Lambda^{(1/D)}_{\alpha\alpha}({\bf k})$ in the expression for the 
compact part of the transverse spin-spin correlation function 
$\tilde\Lambda_{\alpha\alpha}({\bf k})$ (\ref{2.4}) represented in 
fig.2 have the following analytical form:
\begin{eqnarray}\label{3.1}
\hat\Lambda^{(1)}_{\alpha\alpha} & = & 
\tilde\Lambda_{\alpha\alpha zz} l_z \equiv
\tilde\Lambda_{\alpha\alpha zz}\frac{1}{2}
\,v_0\!\!\!\int\!\!\!\frac{d{\bf q}}{(2\pi)^d}
\beta\tilde J_{z\bf q}                                      
\nonumber\\
\hat\Lambda^{(2)}_{\alpha\alpha} & = &
\frac{1}{2}\tilde\Lambda_{\alpha\alpha\beta\beta\gamma\gamma}
\tilde\Lambda_{\beta\beta\gamma\gamma}
\,v_0\!\!\!\int\!\!\!\frac{d{\bf q}}{(2\pi)^d}
V_{\bf q}\tilde V_{\bf q}         \nonumber 
\\
\hat\Lambda^{(3)}_{\alpha\alpha} & = &
\tilde\Lambda_{\alpha\alpha\beta\beta z}\tilde\Lambda_{\beta\beta z}
\,v_0\!\!\!\int\!\!\!\frac{d{\bf q}}{(2\pi)^d}
\beta\tilde J_{z\bf q}\tilde V_{\bf q}   \nonumber 
\\
\hat\Lambda^{(4)}_{\alpha\alpha} & = &
\frac{1}{2}\tilde\Lambda_{\alpha\alpha\beta\beta\gamma\gamma}
\tilde\Lambda_{\beta\beta z}\tilde\Lambda_{\gamma\gamma z}
\,v_0\!\!\!\int\!\!\!\frac{d{\bf q}}{(2\pi)^d}
\beta\tilde J_{z\bf q}\tilde V^2_{\bf q}               
\\
\hat\Lambda^{(5)}_{\alpha\alpha}({\bf k}) & = &
\tilde\Lambda^2_{\alpha\alpha\beta\beta}
\,v_0\!\!\!\int\!\!\!\frac{d{\bf q}}{(2\pi)^d}
\beta\tilde J_{\bf k-q}\tilde V_{\bf q}   \nonumber 
\\
\hat\Lambda^{(6)}_{\alpha\alpha}({\bf k}) & = &
\tilde\Lambda^2_{\alpha\alpha z}
\,v_0\!\!\!\int\!\!\!\frac{d{\bf q}}{(2\pi)^d}
\beta\tilde J_{\bf k-q}\beta\tilde J_{z\bf q} 
\nonumber 
\\
\hat\Lambda^{(7+7')}_{\alpha\alpha}({\bf k}) & = &
2\tilde\Lambda_{\alpha\alpha\beta\beta}
\tilde\Lambda_{\alpha\alpha z}\tilde\Lambda_{\beta\beta z}
\,v_0\!\!\!\int\!\!\!\frac{d{\bf q}}{(2\pi)^d}
\beta\tilde J_{\bf k-q}\beta\tilde J_{z\bf q}\tilde 
V_{\bf q}                                    \nonumber 
\\
\hat\Lambda^{(8)}_{\alpha\alpha}({\bf k}) & = &
\tilde\Lambda_{\alpha\alpha\beta\beta}
\tilde\Lambda_{\alpha\alpha\gamma\gamma}
\tilde\Lambda_{\beta\beta z}\tilde\Lambda_{\gamma\gamma z}
\,v_0\!\!\!\int\!\!\!\frac{d{\bf q}}{(2\pi)^d}
\beta\tilde J_{\bf k-q}\beta\tilde J_{z\bf q}\tilde 
V^2_{\bf q}                                   \nonumber
\end{eqnarray}
Here $\tilde\Lambda_{\alpha\alpha\beta\beta}=
\tilde\Lambda_{\beta\beta\gamma\gamma}$,
$\tilde\Lambda_{\alpha\alpha z}=\tilde\Lambda_{\beta\beta z}$, etc., are 
the renormalized multi-spin cumulants with $\alpha\not=\beta\not=\gamma$ 
(no summation over $\beta$ and $\gamma$ in (\ref{3.1})) given by the 
formulas analogous to (\ref{2.5}). As the diagrams (\ref{3.1}) should be 
calculated only in the \mbox{1-st} 
nonvanishing order in $1/D$, one can use for 
the renormalized transverse interaction line 
$\beta\tilde J_{\bf q}$ 
(see fig.1,c) the simplified expression $\beta\tilde J_{\bf 
q}\cong \beta J_{\bf q}/(1-\tilde\Lambda_{\alpha\alpha}\beta 
J_{\bf q})$, where 
$\tilde\Lambda_{\alpha\alpha} = (\theta/D)G$ and $G$ corresponds to the 
spherical model (see (\ref{2.20})). The renormalized longitudinal 
interaction $\beta\tilde J_{z\bf q}$ (see fig.2,b) is given by
\begin{equation}\label{3.2}
\beta\tilde J_{z\bf q} = \frac{\beta J_{\bf q}}
{1-(\tilde\Lambda_{zz}+\tilde\Lambda^2_{\alpha\alpha z}
\tilde V_{\bf q}) \beta J_{\bf q} }
\end{equation}
and the renormalized 4-spin correlation line $\tilde V_{\bf q}$ 
(see 
fig.2,c) reads $\tilde V_{\bf q}=
V_{\bf 
q}/(1-\tilde\Lambda_{\alpha\alpha\beta\beta}V_{\bf q})$, where
\begin{equation}\label{3.3}
V_{\bf q} = \frac{D}{2}\,v_0\!\!\!\int\!\!\!\frac{d{\bf p}}{(2\pi)^d}
\beta\tilde J_{\bf p}\beta\tilde J_{\bf q-p}
\end{equation}
is the unrenormalized 4-spin correlation line, the factor $D$
(or $D-1$, which plays no role here) in 
(\ref{3.3}) resulting from the summation over the spin-component indices 
$\beta$ and $\gamma$ in the diagrams.
Calculating now the renormalized cumulants $\tilde\Lambda$ in 
(\ref{3.1}) in the lowest order in $1/D$ by the pass method, one gets 
\cite{garst94}
\begin{equation}\label{3.4}
\tilde\Lambda_{\alpha\alpha\beta\beta} \cong
-\left(\frac{2}{D}\right)^3 \frac{1}{(2y)^2(2y-1)}; \qquad
\tilde\Lambda_{\alpha\alpha\beta\beta\gamma\gamma} \cong
\left(\frac{2}{D}\right)^5 \frac{2(3y-1)}{(2y)^3(2y-1)^3}
\end{equation}
with $y\equiv 1/(\theta G)$ and, additionaly, 
\begin{equation}\label{3.5}
\setlength{\arraycolsep}{0.5mm}
\begin{array}{llllll}
\tilde\Lambda_{zz} & \cong & \tilde\Lambda_{\alpha\alpha} + 
\tilde\Lambda_{\alpha\alpha\beta\beta}\xi^2 ; 
&
\qquad
\tilde\Lambda_{\alpha\alpha z} & \cong &
\tilde\Lambda_{\alpha\alpha\beta\beta}\xi               
\\
\tilde\Lambda_{\alpha\alpha zz} & \cong &
\tilde\Lambda_{\alpha\alpha\beta\beta} + 
\tilde\Lambda_{\alpha\alpha\beta\beta\gamma\gamma}\xi^2 ;  
&
\qquad
\tilde\Lambda_{\alpha\alpha\beta\beta z} & \cong &
\tilde\Lambda_{\alpha\alpha\beta\beta\gamma\gamma}\xi     
\end{array}
\end{equation}
where $\xi = (D/\theta)(h+\nu m_0) = Dym_0$. 
With the use of these results for the corresponding contributions 
into $\Delta^{(1/D)}_{\bf k} \equiv 
\lim_{D\to\infty} \left[(D^2/\theta)
\hat\Lambda^{(1/D)}_{\alpha\alpha}({\bf k})\right]$, one obtains
\begin{eqnarray}\label{3.6}
\Delta^{(1)}_{\alpha\alpha} & = & 
-\frac{\nu G^2}{(2y-1)} \left[ 1 - 4m^2_0y\frac{3y-1}{(2y-1)^2} \right]
\,v_0\!\!\!\int\!\!\!\frac{d{\bf q}}{(2\pi)^d}
\tilde\lambda_{z\bf q}                                      
\nonumber\\
\Delta^{(2)}_{\alpha\alpha} & = &
-2G^5\frac{3y-1}{(2y-1)^4}
\,v_0\!\!\!\int\!\!\!\frac{d{\bf q}}{(2\pi)^d}
\varphi_{\bf q}\tilde\varphi_{\bf q}                   
     \nonumber\\
\Delta^{(3)}_{\alpha\alpha} & = &
-8\nu m^2_0yG^4\frac{3y-1}{(2y-1)^4}
\,v_0\!\!\!\int\!\!\!\frac{d{\bf q}}{(2\pi)^d}
\tilde\lambda_{z\bf q}\tilde\varphi_{\bf q}            
          \\
\Delta^{(4)}_{\alpha\alpha} & = &
4\nu m^2_0yG^6\frac{3y-1}{(2y-1)^5}
\,v_0\!\!\!\int\!\!\!\frac{d{\bf q}}{(2\pi)^d}
\tilde\lambda_{z\bf q}\tilde\varphi^2_{\bf q}          
     \nonumber\\
\Delta^{(5)}_{\alpha\alpha}({\bf k}) & = &
\frac{2\nu G^4}{(2y-1)}
\,v_0\!\!\!\int\!\!\!\frac{d{\bf q}}{(2\pi)^d}
\tilde\lambda_{\bf k-q}\tilde\varphi_{\bf q}           
     \nonumber\\
\Delta^{(6+7+7'+8)}_{\alpha\alpha}({\bf k}) & = &
\frac{4m^2_0yG^3}{(2y-1)^2}
\,v_0\!\!\!\int\!\!\!\frac{d{\bf q}}{(2\pi)^d}
\tilde\lambda_{\bf k-q}
\tilde\lambda_{z\bf q} L^2_{\bf q}    \nonumber
\end{eqnarray}
where $\tilde\varphi_{\bf q} \equiv \varphi_{\bf 
q}L_{\bf q} \equiv 
\varphi_{\bf q}/(1+a\varphi_{\bf q})$, $a \equiv 
G^2/(2y-1)$,
\begin{equation}\label{3.7}
\varphi_{\bf q} = 
\,v_0\!\!\!\int\!\!\!\frac{d{\bf p}}{(2\pi)^d}
\tilde\lambda_{\bf p}
\tilde\lambda_{\bf p-q}; \qquad
\tilde\lambda_{\bf q} \equiv 
\tilde J_{\bf q}/J_0  \equiv
\frac{\lambda_{\bf q}}
{1-\nu G\lambda_{\bf q}}
\end{equation}
The expression for the renormalized $z$-interaction line 
$\tilde\lambda_{z\bf q} 
\equiv \tilde J_{z\bf q}/J_0$  can be written in the form
\begin{equation}\label{3.8}
\tilde\lambda_{z\bf q} = 
\frac{\lambda_{\bf q}}{1-\nu G_{z\bf q} 
\lambda_{\bf q}} ; \qquad
G_{z\bf q} = G 
\left(
1 - \frac{2m^2_0y}{2y-1}L_{\bf q} 
\right)
\end{equation}
Further simplifications leading to the results listed at the end of the 
Section 2 can be archieved if one expresses 
$\varphi_{\bf q}$ through $r_{\bf q}$ 
(\ref{3.9}) with the use of the relation 
$(2y-1)(1+a\varphi_{\bf q}) = r_{\bf q} + 
2m^2_0y$.

\end{document}